\documentclass[prl,twocolumn,showpacs,amsmath,amssymb]{revtex4}
\usepackage{graphicx}
\usepackage{bm}
\usepackage{amsmath,amssymb}

\begin{document}

\title{How to upload a physical state to the correlation space} 

\author{Tomoyuki Morimae}
\email{morimae@gmail.com}
           \affiliation{
Laboratoire Paul Painlev\'e, Universit\'e Lille 1,
F-59655 Villeneuve d'Ascq Cedex, France
}
\date{\today}
            
\begin{abstract}
In the framework of the computational tensor network
[D. Gross and J. Eisert, Phys. Rev. Lett. {\bf98}, 220503 (2007)], 
the quantum computation is performed in a virtual linear space
which is called the correlation space. 
It was recently shown 
[J. M. Cai, W, D\"ur, M. Van den Nest, A. Miyake, and H. J. Briegel,
Phys. Rev. Lett. {\bf103}, 050503 (2009)] 
that a state in the correlation
space can be downloaded to the real physical space.
In this letter, conversely, we 
study how to upload a state from a real physical space to
the correlation space being motivated by the virtual-real hybrid
quantum information processing. After showing the impossibility of the cloning
of a state between the real physical space and the correlation space,
we propose a simple teleportation-like method of the upload.
Applications of this method 
also enable the Gottesman-Chuang gate teleportation trick and the entanglement
swapping in the virtual-real hybrid setting.
Furthermore, compared with the inverse of the downloading method by Cai, et. al.,
which also works as
the upload, our uploading method has several advantages.
\end{abstract}
\maketitle  
{\it Introduction}.---
Quantum many-body states, which have long been the central research
objects in condensed
matter physics, statistical physics, quantum chemistry, and nuclear physics,
are now receiving a renewed interest in quantum information
science as a fundamental resource for the quantum computation.
One canonical example of such a resource state is the
cluster state~\cite{one-way};
once the cluster state is prepared,
the universal quantum computation is possible
with the adaptive measurements on each qubit. 
Recently, 
the framework of the computational tensor
network was proposed~\cite{Gross},
which enables us to have a bird's-eye-view in the exploration
of many-body resource states beyond the cluster state.
The most exciting feature of this framework is the clever use of
{\it the correlation space}, which
is a virtual linear space where the quantum computation is performed.
A virtual state which lives in the correlation space 
is synchronized with a set of real
physical qubits, and the universal quantum operations,
including the initialization and the measurement,
on the virtual state is driven by projection
measurements on these physical qubits.
Since the way of the synchronization 
is determined by how these physical qubits
are entangled with each other, 
the framework of the computational tensor network
offers the fresh motivation for studying
the multipartite entanglement in quantum many-body Hamiltonians.

\begin{figure}[htbp]
\begin{center}
\includegraphics[width=0.4\textwidth]{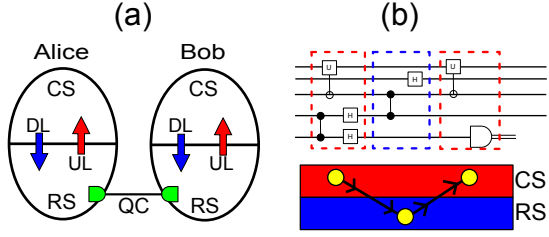}
\end{center}
\caption{(Color online.) 
(a): Quantum computation on distributed correlation spaces.
Alice who has her own correlation space (CS)
downloads (DL) a state from her correlation space
to her real physical space (RS),
and sends it to Bob through the quantum channel (QC).
Bob who receives real physical qubits from Alice uploads (UL) them
to his own correlation space for his calculation.
(b): The strategy of avoiding less stable space in the virtual-real
hybrid QC. The region 
surrounded by the red (blue) dotted box
is the one where the correlation space (real space) is more stable than
the other. The yellow circle represents the quantum register.
When the correlation (real) space is more stable than the other,
the register stays there.
Once the space where the register lives becomes less stable than
the other, the register emigrates.
} 
\label{escape}
\end{figure}

Although the 
computational tensor network is sufficient for the universal quantum 
computation, however, 
it does not provide any method for 
the download and the upload of {\it a quantum state} 
between the real physical space and the correlation space.
There are, and will be, several situations where such a method is needed.
For example,
the most important one would be the {\it virtual-real 
hybrid quantum information processing}: 
if we want to perform the quantum computation on
distributed correlation spaces (Fig.~\ref{escape} (a)),
methods for the download and the upload of quantum states 
are indispensable.
Furthermore, since noises and errors 
in the correlation space are not necessarily the same as
those in the corresponding real physical space, we
might be able to increase the stability of the quantum computation by taking the
strategy of avoiding
the less stable space (Fig.~\ref{escape} (b)).
Recently, a method for the download of a state from the correlation space
to the real physical space was proposed
in Ref.~\cite{Cai}.
This breakthrough has made the framework of the computational tensor network
the universal state preparator~\cite{power1}.

\begin{figure}[htbp]
\begin{center}
\includegraphics[width=0.4\textwidth]{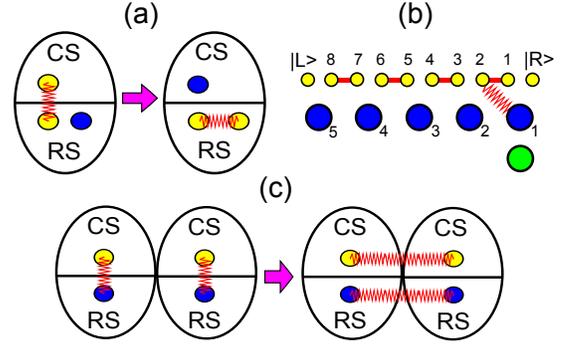}
\end{center}
\caption{(Color online.) (a): The interpretation of our uploading
method in terms of the quantum teleportation.
The left of (a): The red zigzag line represents the 
entanglement-like correlation between the correlation space (CS) and
the real space (RS). 
The right of (a): After the projection measurement in the 
Bell basis on the two qubits in
the real space, the blue state is uploaded (teleported) to the correlation
space. The red zigzag line represents the real entanglement.
(b): The interpretation of the same uploading method as (a)
in terms of the 
valence-bond (VB) 
picture~\cite{RVB}. 
The red zigzag line represents the entanglement-like correlation.
The Bell measurement
is performed on the first big blue qubit and the big green qubit.
The state of the big green qubit is ``teleported" to the second yellow
small qubit.
(c): The entanglement swapping (repeating). 
The Bell measurement on the two blue qubits in real spaces
creates the entanglement between
the two yellow qubits in correlation spaces.
} 
\label{teleportation}
\end{figure}

In this letter, conversely, we propose a method for uploading a state from
the real physical space to the correlation space,
which completes the needed I/O infrastructure  
for the virtual-real hybrid quantum information processing.
We first show the impossibility of the cloning of a state between the correlation
space and the real physical space. We next point out several
disadvantages 
of using the inverse of 
the downloading method of Ref.~\cite{Cai} for the purpose of the upload.
We then explain our uploading method,
which is simple and free from those disadvantages.
Interestingly, our uploading method 
is interpreted as
a ``teleportation" from the real physical space to the correlation
space which 
consumes an entanglement-like correlation between the real physical
space and the correlation
space (Fig.~\ref{teleportation} (a) and (b)):
if this entanglement-like correlation
is maximum, 
the fidelity of the upload is unity,
whereas if it is not, the upload succeeds only
probabilistically.
Furthermore, our uploading method
also enables several important tricks of quantum information processings,
such as the Gottesman-Chuang gate teleportation trick~\cite{teleQC}
and the entanglement swapping (Fig.~\ref{teleportation} (c)) 
in the virtual-real hybrid setting.

{\it Downloading method}.---
The framework of the computational tensor 
network~\cite{Gross} starts with the 
matrix product state 
$\Phi\big(|R\rangle\big)_1^N\equiv\sum_{l_1,...,l_N}
\langle L|A[l_N]...A[l_1]|R\rangle
|l_N,...,l_1\rangle$
of $N$ qubits,
where $l_i\in \{0,1\}$ ($i=1,...,N$), 
$|L\rangle$ and $|R\rangle$ are two-dimensional complex vectors,
$A[0]$ and $A[1]$ are $2\times2$ complex matrices,
and $|l_i\rangle$ is the state of $i$th physical qubit
in the computational basis $\{|0\rangle,|1\rangle\}$.
The quantum computation is performed not in the Hilbert space spanned
by $|l_i\rangle$'s but in the linear space, which is called
the correlation space, where
$|L\rangle$, $|R\rangle$, $A[0]$, and $A[1]$ live.

It was shown~\cite{Cai} that a state in the correlation
space can be downloaded to the real physical space.
Assume that
there exists a local basis $\{|m_0\rangle,|m_1\rangle\}$
such that
\begin{eqnarray}
A[m_0]=|\phi_0\rangle\langle0|,~~
A[m_1]=|\phi_1\rangle\langle1|
\label{Asimple}
\end{eqnarray}
with $\langle\phi_i|\phi_j\rangle=\delta_{i,j}$,
where
$A[m_s]\equiv\langle m_s|0\rangle A[0]
+\langle m_s|1\rangle A[1]$.
Then, by starting with $\Phi(|\psi\rangle)_1^N$,
where $|\psi\rangle$ is the 
state which we want to download,
the method of Ref.~\cite{Cai} gives the state
\begin{eqnarray}
\sum_{s=0}^1\Phi(|\phi_s\rangle)_{k+1}^N\otimes|m_s\rangle_k\otimes
(\tilde{Z}^s|\tilde{\psi}\rangle_1),
\label{Caieq}
\end{eqnarray}
where $\tilde{Z}$ and $|\tilde{\psi}\rangle$
are $Z$ and $|\psi\rangle$
in terms of the $\{|m_s\rangle\}$ basis.
The projection measurement on the $k$th qubit 
in the $\{|m_s\rangle\}$ basis
completes the desired download up to the phase error 
$\tilde{Z}^s$ and the change of the basis
$\{|0\rangle,|1\rangle\}\to\{|m_0\rangle,|m_1\rangle\}$.
In general, $A[0]$ and $A[1]$ have more complicated 
structures than Eq.~(\ref{Asimple}).
According to Ref.~\cite{Gross}, the most general
form is
\begin{eqnarray}
A[0]\propto W,~~
A[1]\propto W\mbox{diag}(e^{-i\alpha},e^{i\alpha})
\label{Acomplicated}
\end{eqnarray}
for some $W\in SU(2)$.
In this case, the filtering~\cite{Cai}
$\{\mathcal{F},\bar{\mathcal{F}}\}$ 
is required 
in order to make some non-orthogonal physical states orthogonal.
The filtering succeeds with a certain
probability, and in this case the download succeeds.
If the filtering fails,
a single physical qubit is consumed. Then, we repeat the filtering
on other remaining physical qubits until we succeed.

\begin{figure}[htbp]
\begin{center}
\includegraphics[width=0.3\textwidth]{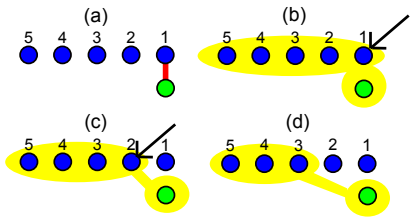}
\end{center}
\caption{(Color online.) 
The inverse of the downloading method of Ref.~\cite{Cai}.
(a): Blue circles are real physical qubits of $\Phi(|+\rangle)_1^N$. 
The green circle is the state
$|\psi\rangle$ which we want to upload. 
The controlled-Z (the red bond) is applied between 
the first physical qubit in $\Phi(|+\rangle)_1^N$ and $|\psi\rangle$.
(b): The first blue qubit is projected.
(c): The second blue qubit is projected. In this way, several blue qubits are
projected.
(d): Finally, the state of qubits that are not measured
(i.e., qubits in the yellow region) is our goal $\Phi(|\psi\rangle)$ 
where $|\psi\rangle$ is uploaded to the correlation space.
Note that if the entanglement between the green qubit and blue qubits
is destroyed in the middle of the process, the upload fails.
} 
\label{Cai_inverse}
\end{figure}

{\it Uploading method}.---
Conversely, how to upload a state from
a real physical space to the correlation space?
One naive way of the upload is to do the inverse of the 
above downloading method
(see Fig.~\ref{Cai_inverse} and Appendix 1 and 2).
However, it is worth exploring other possibilities
since the inverse of the above downloading method
has several disadvantages: Firstly, the inverse is
complicated even when $A[0]$ and $A[1]$ have the simple form 
Eq.~(\ref{Asimple}) (see Appendix 1).
Secondly, if $A[0]$ and $A[1]$ take the 
most general form Eq.~(\ref{Acomplicated}),
we must invert the filtering process $\{\mathcal{F},\bar{\mathcal{F}}\}$.
Although the inversion of the filtering 
$\{\mathcal{F},\bar{\mathcal{F}}\}$
is in principle possible by introducing
the new filtering
$\{\mathcal{G},\bar{\mathcal{G}}\}$ (see Appendix 2),
it is cumbersome to prepare two different filterings for 
the download and the upload. 
Finally, and most crucially, we must  
maintain the entanglement between the state $|\psi\rangle$ which we want to upload
and the place $\Phi$ where we want to upload
during the whole process of the inversion (see Fig.~\ref{Cai_inverse}):
if the entanglement between $|\psi\rangle$ and $\Phi$
is destroyed in the middle of the inversion process, the upload fails
and, what is worse, the original state is destroyed
(see Appendix 1).
This disadvantage makes it difficult to upload a state of, for example, a
photon, which is hard to be localized,
to the correlation space of stationary qubits, such as atoms, ions, and quantum
dots.
As we will see later, our uploading method is free from those
three disadvantages.

The other possibility for the way of the upload might be the cloning~\cite{clone} 
of a state from the real physical
space to the correlation space. However, 
we can see that it is impossible.
Assume that we can clone 
any state $|\psi\rangle$ from the real physical space to the correlation
space:
$\Phi(|e\rangle)\otimes|\psi\rangle
\to\Phi(|\psi\rangle)\otimes|\psi\rangle$,
where $|e\rangle$ is a fixed virtual initial state.
Then, if we prepare infinitely many $\Phi(|e\rangle)$'s and 
clone $|\psi\rangle$ to each $\Phi(|e\rangle)$, 
we have infinitely many $\Phi(|\psi\rangle)$'s,
from which we can know the shape of $|\psi\rangle$
without destroying the original.
(For a similar reason, the cloning 
$\Phi(|\psi\rangle)\otimes|e\rangle\to\Phi(|\psi\rangle)\otimes|\psi\rangle$
from the correlation space to
the real physical space is also impossible. If it is possible, 
we can first upload a real physical state $|\psi\rangle$ to
the correlation space by using our uploading method which will be
explained later,
and then repeat the cloning
from the correlation space to the real space, which 
provides an infinitely many copies of $|\psi\rangle$ in the real physical space.
Indeed, we can easily verify from Eq.~(\ref{Caieq}) that
the downloading method
of Ref.~\cite{Cai} destroys the original in the correlation space.)

Since the cloning is impossible, a compromise is the quantum 
teleportation~\cite{tele}, which regrettably destroys the original state.
Indeed, a teleportation-like method of the 
upload from the real physical space to
the correlation space is possible.
Let us start with the state 
$\Phi(|R\rangle)_1^N\otimes|\psi\rangle_0$,
where  
$|\psi\rangle=\lambda_0|0\rangle+\lambda_1|1\rangle$ 
is the state which we want to upload
and the subscript 0 of $|\psi\rangle_0$ indicates that
$|\psi\rangle$ is placed on the zeroth site.
We can take any state as $|R\rangle$
by running an appropriate pre-processing. 
By a straightforward calculation,
$\Phi(|R\rangle)_1^N\otimes|\psi\rangle_0
=\sum_{s,r=0}^1\lambda_r\Phi(A[m_s]|R\rangle)_2^N
\otimes|m_s\rangle_1\otimes|r\rangle_0$,
where $\{|m_0\rangle,|m_1\rangle\}$ is a certain orthonormal basis,
and $A[m_s]\equiv\langle m_s|0\rangle A[0]+\langle m_s|1\rangle A[1]$.
Let us project the zeroth and first qubits onto
$|B_1\rangle_{10}\equiv|m_0\rangle_1\otimes|0\rangle_0
+|m_1\rangle_1\otimes|1\rangle_0$.
Then, we obtain
$\Phi(\lambda_0A[m_0]|R\rangle+\lambda_1A[m_1]|R\rangle)_2^N
\otimes|B_1\rangle_{10}$.
If $A[m_0]|R\rangle$ and $A[m_1]|R\rangle$
are orthogonal with each other, the upload is
succeeded 
up to the trivial change of the basis 
$\{|0\rangle,|1\rangle\}\to
\{A[m_0]|R\rangle,A[m_1]|R\rangle\}$, which is compensated
by a post-processing.
For example, the orthogonality between 
$A[m_0]|R\rangle$ and $A[m_1]|R\rangle$ is satisfied
for the one-dimensional cluster state~\cite{one-way}
whose
matrix product state is given by~\cite{Gross}
$A[0]=|+\rangle\langle0|$
and
$A[1]=|-\rangle\langle1|$.
 
It is not difficult to see the strong similarity of the above uploading method
to the quantum teleportation~\cite{tele} if
we interpret
$A[m_s]|R\rangle$ (not $\Phi(A[m_s]|R\rangle)$!) and $|m_s\rangle_1$ in
\begin{eqnarray}
\sum_{s=0}^1\Phi(A[m_s]|R\rangle)_2^N\otimes|m_s\rangle_1
\label{entangled}
\end{eqnarray}
as being ``entangled" with each other (see Fig.~\ref{teleportation} (a) and (b)).
Although this entanglement-like correlation is not always connected
to the real physical entanglement between the first qubit and other qubits,
since the orthogonality of $A[m_0]|R\rangle$
and $A[m_1]|R\rangle$ 
does not necessarily mean that of 
$\Phi(A[m_0]|R\rangle)_2^N$
and
$\Phi(A[m_1]|R\rangle)_2^N$ (see Appendix 3),
this analogy (and, after all, the mathematical equivalency)
between our uploading method and 
the quantum teleportation
is fascinating.
For example,
the analogy makes it easy to understand what happens if the
zeroth and first qubits are projected onto not $|B_1\rangle$ 
but one of other Bell
basis
$|B_2\rangle\equiv|m_0,0\rangle-|m_1,1\rangle$,
$|B_3\rangle\equiv|m_0,1\rangle+|m_1,0\rangle$,
and
$|B_4\rangle\equiv|m_0,1\rangle-|m_1,0\rangle$
in our uploading method.
In this case, as the analogy indicates, 
the uploaded state is affected by some errors
depending on the measurement result.
If $A[m_0]|R\rangle$ and $A[m_1]|R\rangle$ are orthogonal with
each other, these errors are the phase error or the bit error
in terms of
the $\{A[m_0]|R\rangle,A[m_1]|R\rangle\}$ basis.
Such an error is corrected by a post-processing,
and therefore 
the fidelity of the upload is unity.
Indeed, if $A[m_0]|R\rangle$ and $A[m_1]|R\rangle$
are orthogonal with each other, 
it is reasonable to think that
the virtual ``entanglement" in Eq.~(\ref{entangled}) 
is ``maximum".

If $A[m_0]|R\rangle$ and $A[m_1]|R\rangle$
are not orthogonal with each other,  
on the other hand,
we are inclined to think that
the virtual ``entanglement" in Eq.~(\ref{entangled})
is not ``maximum". Then,
the analogy with the quantum teleportation
indicates that the upload succeeds only
partially.
Indeed, it is the case.
If $A[m_0]|R\rangle$ and $A[m_1]|R\rangle$
are not orthogonal with each other, the uploaded state
$\lambda_0A[m_0]|R\rangle+\lambda_1A[m_1]|R\rangle$
is distorted from the original,
and therefore the fidelity of the upload
is less than unity.
In this case, by applying
the same filtering $\{\mathcal{F},\bar{\mathcal{F}}\}$ of Ref.~\cite{Cai},
we can perform a perfect but probabilistic
upload. 
Let us assume that $A[0]$ and $A[1]$ take the most general form Eq.~(\ref{Acomplicated}).
Then, there always exists 
a basis $\{|m_0\rangle,|m_1\rangle\}$
such that
$A[m_0]=r_0|\phi_0\rangle\langle0|$ and
$A[m_1]=r_1|\phi_0\rangle\langle0|
+|\phi_1\rangle\langle1|$,
where $\langle\phi_i|\phi_j\rangle=\delta_{i,j}$, 
$r_0>0$, $r_1\ge0$, and $r_0^2+r_1^2=1$~\cite{Cai}.
By rewriting $\Phi(|+\rangle)_1^N$ as
$\sum_{s=0}^1\Phi(|\phi_s\rangle)_2^N\otimes|m_s'\rangle_1$,
where $|m_0'\rangle\equiv r_0|m_0\rangle+r_1|m_1\rangle$
and $|m_1'\rangle\equiv|m_1\rangle$,
and applying the filtering operation $\{\mathcal{F},\bar{\mathcal{F}}\}$ 
(for the definition of it, see Appendix 2 or Ref.~\cite{Cai})
on the first qubit, we probabilistically obtain
$\sum_{s=0}^1\Phi(|\phi_s\rangle)_2^N\otimes|m_s\rangle_1$
if $\mathcal{F}$ is realized.
In this way, we obtain the state which has
the ``maximum" entanglement-like correlation between the
real physical space and the correlation space.
With this state,
we can deterministically upload a physical state to 
the correlation space by using our teleportation-like method 
explained previously.
If the filtering fails (i.e., if
$\bar{\mathcal{F}}$ is realized), the net effect of
the filtering 
is the consumption of the first qubit.
Therefore, 
we have only to repeat the same filtering
on the remaining qubits
until we succeed. 

{\it Discussion}.---
In this letter, we have proposed the 
teleportation-like method of the upload of a state
from a real physical space to the correlation space.
Our method is free from
the disadvantages of the inverse of the downloading method of Ref.~\cite{Cai}:
Firstly, our method is simpler and more intuitive. Secondly, 
no new filtering is needed for the upload; 
once the filtering 
$\{\mathcal{F},\bar{\mathcal{F}}\}$
is prepared, the download is possible with the method of Ref.~\cite{Cai}
and the upload is possible with our method.
Finally, our method does not require any long-time interaction between
the state $|\psi\rangle$ which we want
to upload and the place $\Phi$ where we want to upload.

The assumption of the availability of the
Bell measurement is, though something to be avoided if possible,
not too demanding, since it is ubiquitous
in many quantum information protocols, such as 
the quantum teleportation~\cite{tele},
the quantum dense coding~\cite{dense},
the quantum repeater~\cite{repeater},
and the teleportation-based quantum computation~\cite{teleQC}.
In fact, if we want to do the upload,
a two-qubit operation is unavoidable:
as long as we start with 
$\Phi(|R\rangle)_1^N\otimes|\psi\rangle_0$,
where $|\psi\rangle$ is the state which we want to upload,
we must entangle the zeroth qubit with at least one physical qubit
in $\Phi(|R\rangle)_1^N$, since otherwise 
$\Phi(|R\rangle)_1^N$ cannot ``know" $|\psi\rangle$.
Among many possibilities which use two-qubit operations, 
the teleportation seems to be the most feasible and the most familiar one,
since many experimental schemes of the teleportation are available in
the optical system, 
the NMR,
the trapped ions,
solids,
and the light-matter hybrid~\cite{Furusawa}.
Note that a universal set of single-qubit rotations, two-qubit entangling
gates, and the Deutsch's algorithm in the correlation space 
have been recently realized~\cite{experiment} with 
the four- and six-qubit optical systems.
In addition to such an optical system, the systems
of bosonic atoms or spin-1 polar molecules trapped in a 
three-dimensional optical lattice, where the AKLT model is realized through
the tunneling-induced collision~\cite{Yip}
or the microwave-induced dipole-dipole interaction~\cite{Brennen}, respectively,
would be promising places where our uploading 
method could be implemented, since in the one-dimensional AKLT chain, 
the right boundary spin-1/2 particle is strongly correlated
with the virtual state in the correlation space~\cite{MiyakeAKLT}. 

As is easily shown, the teleportation-like method for the
download is also possible.
However, it will not offer any advantage over
the original downloading method of Ref.~\cite{Cai},
since we must implement the Bell measurement in the correlation space,
which needs the coupling of two computational quantum wires.



Applications of our uploading method enable 
several important tricks of quantum information protocols
in the virtual-real hybrid setting.
For example, we can perform the entanglement swapping
between the real spaces and the correlation spaces 
as is illustrated in Fig.~\ref{teleportation} (c)
(see Appendix 5). Such a swapping method 
is useful to establish the entanglement
between two correlation spaces without touching them.
We can also perform the 
Gottesman-Chuang gate teleportation trick~\cite{teleQC} 
from the real space to the correlation space (see Appendix 4); 
if there are some gates which are difficult
to be implemented in the correlation spaces, we can prepare these gates offline
in the real spaces and ``wedge" them into the correlation spaces
as in Ref.~\cite{KLM}.
In summary, we believe that our uploading method in addition to the downloading method of 
Ref.~\cite{Cai} open the door to the new framework of the quantum information
processing, namely the virtual-real hybrid quantum information processing.

The author is supported by the 
ANR (StatQuant, JC07 07205763).


\newpage


{\bf Appendix 1: The inverse of the downloading method of Ref.~\cite{Cai}}.---
Let us first consider the case of Eq.~(\ref{Asimple}).
We start with 
\begin{eqnarray*}
\Phi(|+\rangle)_1^N=\sum_{s=0}^1\Phi(|\phi_s\rangle)_2^N\otimes|m_s\rangle_1.
\end{eqnarray*}
Let us couple it with the state 
\begin{eqnarray*}
|\tilde{\psi}\rangle\equiv\lambda_0|m_0\rangle
+\lambda_1|m_1\rangle,
\end{eqnarray*}
which is the one we want to upload in terms of
the $\{|m_0\rangle,|m_1\rangle\}$ basis, and apply the
controlled-$Z$ gate (in terms of the $\{|m_0\rangle,|m_1\rangle\}$ basis) 
between the first qubit and $|\tilde{\psi}\rangle$: 
\begin{eqnarray*}
\sum_{s=0}^1\Phi(|\phi_s\rangle)_2^N\otimes|m_s\rangle_1\otimes
(\tilde{Z}^s|\tilde{\psi}\rangle_0),
\end{eqnarray*}
where 
\begin{eqnarray*}
\tilde{Z}\equiv |m_0\rangle\langle m_0|
-|m_1\rangle\langle m_1|
\end{eqnarray*}
is the Pauli $z$ operator in terms of the 
$\{|m_0\rangle,|m_1\rangle\}$ 
basis, and
$|\tilde{\psi\rangle}$ is placed on the zeroth site.
It is rewritten as
\begin{eqnarray*}
\lambda_0\Phi(|+\rangle)_1^N\otimes|m_0\rangle_0
+\lambda_1\Phi(|-\rangle)_1^N\otimes|m_1\rangle_0.
\end{eqnarray*}
Note that if the zeroth site is measured in the 
$\{|m_0\rangle,|m_1\rangle\}$ basis
at this stage, the entanglement between the zeroth site and other sites
is destroyed, and therefore the information about $\{\lambda_0,\lambda_1\}$ 
is lost. Then, the upload fails.

By implementing the rotation 
$\{|+\rangle,|-\rangle\}\to\{|\phi_0\rangle,|\phi_1\rangle\}$
in the correlation space,
which is always possible by the definition of the computational quantum
wire,
we obtain
\begin{eqnarray*}
\lambda_0\Phi(|\phi_0\rangle)_k^N\otimes|m_0\rangle_0
+\lambda_1\Phi(|\phi_1\rangle)_k^N\otimes|m_1\rangle_0
\end{eqnarray*}
for certain $k$. Again, if the entanglement between the zeroth site
and other sites is destroyed at this stage, the upload fails.
It is rewritten as
\begin{eqnarray*}
\sum_{s=0}^1\Phi(A[m_s]|\psi\rangle)_k^N\otimes|m_s\rangle_0.
\end{eqnarray*}
By changing the basis of the zeroth site to the computational basis, 
and swapping
the zeroth site with the $k-1$th site,
we obtain
$\Phi(|\psi\rangle)_{k-1}^N$,
which is our goal. 
Note that the original state (zeroth site)
is interacting with $\Phi$ from the beginning to the end.


{\bf Appendix 2: Filtering}.---
Let us next consider the case of Eq.~(\ref{Acomplicated}).
In this case, there always exists 
a basis $\{|m_0\rangle,|m_1\rangle\}$
such that
\begin{eqnarray*}
A[m_0]=r_0|\phi_0\rangle\langle0|,~~
A[m_1]=r_1|\phi_0\rangle\langle0|
+|\phi_1\rangle\langle1|,
\end{eqnarray*}
where $\langle\phi_i|\phi_j\rangle=\delta_{i,j}$, 
$r_0>0$, $r_1\ge0$, and $r_0^2+r_1^2=1$~\cite{Cai}.
Let us define
\begin{eqnarray*}
|m_0'\rangle&\equiv&r_0|m_0\rangle+r_1|m_1\rangle\\
|m_1'\rangle&\equiv&|m_1\rangle.
\end{eqnarray*}

In principle, the download for the case of Eq.~(\ref{Acomplicated})
is performed in a similar way as that
for the case of Eq.~(\ref{Asimple}).
However, in the case of Eq.~(\ref{Acomplicated}), the filtering~\cite{Cai}
$\{\mathcal{F},\bar{\mathcal{F}}\}$ 
is required 
in order to transform non-orthogonal states $\{|m_0'\rangle,|m_1'\rangle\}$ 
into orthogonal states $\{|m_0\rangle,|m_1\rangle\}$.
Here,
\begin{eqnarray*}
\mathcal{F}^\dagger\mathcal{F}
+\bar{\mathcal{F}}^\dagger\bar{\mathcal{F}}=I,
\end{eqnarray*}
\begin{eqnarray*}
\mathcal{F}&\equiv&\frac{1}{\sqrt{1+r_1}}
\Big(|m_0\rangle\langle m_0|+r_0|m_1\rangle\langle m_1|
-r_1|m_1\rangle\langle m_0|\Big),\\
\bar{\mathcal{F}}&\equiv&\sqrt{\frac{2r_1}{1+r_1}}|\xi\rangle\langle\xi|,
\end{eqnarray*}
and
\begin{eqnarray*}
|\xi\rangle\equiv\sqrt{\frac{1-r_1}{2}}|m_0\rangle
+\sqrt{\frac{1+r_1}{2}}|m_1\rangle.
\end{eqnarray*}

Let us consider how to invert the downloading method of Ref.~\cite{Cai}
when the filtering is required. We start with
\begin{eqnarray*}
\Phi(|+\rangle)_1^N=\sum_{s=0}^1\Phi(|\phi_s\rangle)_2^N\otimes 
|m_s'\rangle_1.
\end{eqnarray*}
By using the filtering $\{\mathcal{F},\bar{\mathcal{F}}\}$,
we can do the transformation
$\{|m_0'\rangle,|m_1'\rangle\}\to\{|m_0\rangle,|m_1\rangle\}$:
\begin{eqnarray*}
\sum_{s=0}^1\Phi(|\phi_s\rangle)_2^N\otimes 
|m_s\rangle_1.
\end{eqnarray*}
Let us implement the controlled-Z gate (in terms of the 
$\{|m_0\rangle,|m_1\rangle\}$ 
basis) between the first qubit and 
\begin{eqnarray*}
|\tilde{\psi}\rangle\equiv\lambda_0|m_0\rangle+\lambda_1|m_1\rangle,
\end{eqnarray*}
which is the state we want to upload in terms of the
$\{|m_0\rangle,|m_1\rangle\}$ basis.
Then,
we obtain
\begin{eqnarray*}
\sum_{s=0}^1\Phi(|\phi_s\rangle)_2^N\otimes 
|m_s\rangle_1\otimes(\tilde{Z}^s|\tilde{\psi}\rangle_0),
\end{eqnarray*}
where $|\psi\rangle$ is placed on the zeroth site,
and $\tilde{Z}$ 
is $Z$ in terms of the $\{|m_0\rangle,|m_1\rangle\}$ basis.
If we can do the transformation 
$\{|m_0\rangle,|m_1\rangle\}\to\{|m_0'\rangle,|m_1'\rangle\}$,
we obtain
\begin{eqnarray*}
\sum_{s=0}^1\Phi(|\phi_s\rangle)_2^N\otimes|m_s'\rangle_1\otimes
(\tilde{Z}^s|\tilde{\psi}\rangle_0).
\end{eqnarray*}
By a streightforward calculation, it is equivalent to
\begin{eqnarray*}
\lambda_0\Phi(|+\rangle)_1^N\otimes |m_0\rangle_0
+\lambda_1\Phi(|-\rangle)_1^N\otimes |m_1\rangle_0.
\end{eqnarray*}
By implementing the rotation 
$\{|+\rangle,|-\rangle\}\to\{|\phi_0\rangle,|\phi_1\rangle\}$
in the correlation space, which is alway possible if $N$ is sufficiently large,
we obtain
\begin{eqnarray*}
\sum_{s=0}^1\lambda_s\Phi(|\phi_s\rangle)_k^N\otimes |m_s\rangle_0
\end{eqnarray*}
for certain $k$.
By using the filtering $\{|m_s\rangle\}\to\{|m_s'\rangle\}$,
we obtain
\begin{eqnarray*}
\sum_{s=0}^1\lambda_s\Phi(|\phi_s\rangle)_k^N\otimes |m_s'\rangle_0
=
\Phi(|\psi\rangle),
\end{eqnarray*}
which is our goal.

The operation
$\{|m_0\rangle,|m_1\rangle\}\to\{|m_0'\rangle,|m_1'\rangle\}$
is possible, for example, with the filtering
$\{\mathcal{G},\bar{\mathcal{G}}\}$, where
\begin{eqnarray*}
\mathcal{G}^\dagger\mathcal{G}
+\bar{\mathcal{G}}^\dagger\bar{\mathcal{G}}
=I,
\end{eqnarray*}
\begin{eqnarray*}
\mathcal{G}&\equiv&\frac{1}{\sqrt{1+r_1}}
\Big(r_0|m_0\rangle\langle m_0|+r_1|m_1\rangle\langle m_0|
+|m_1\rangle\langle m_1|\Big)\\ 
\bar{\mathcal{G}}&\equiv&
\sqrt{\frac{2r_1}{1+r_1}}|\eta\rangle\langle\eta|,
\end{eqnarray*}
and
\begin{eqnarray*}
|\eta\rangle\equiv\frac{1}{\sqrt{2}}(|m_0\rangle-|m_1\rangle).
\end{eqnarray*}


{\bf Appendix 3: Calculation of the inner product}.---
Let us calculate the inner product between 
$\Phi(|+\rangle)_1^N$
and
$\Phi(|-\rangle)_1^N$
with
$A[0]=|\phi_0\rangle\langle0|$,
$A[1]=|\phi_1\rangle\langle1|$,
and $|L\rangle=|0\rangle$,
where
\begin{eqnarray*}
|\phi_0\rangle&\equiv&\cos\frac{\theta}{2}|0\rangle
+\sin\frac{\theta}{2}|1\rangle\\
|\phi_1\rangle&\equiv&\sin\frac{\theta}{2}|0\rangle
-\cos\frac{\theta}{2}|1\rangle.
\end{eqnarray*}
By the contraction, the inner product is
\begin{eqnarray*}
\sum_{l_N,...,l_1=0}^1
\langle-|A[l_1]^\dagger...A[l_N]^\dagger|0\rangle\langle0|
A[l_N]...A[l_1]|+\rangle,
\end{eqnarray*}
which is interpreted as
\begin{eqnarray*}
\langle-|\Big[
\mathcal{E}\circ...\circ\mathcal{E}(|0\rangle\langle0|)\Big]|+\rangle
\end{eqnarray*}
with the map
\begin{eqnarray*}
\mathcal{E}(\rho)\equiv\sum_{s=0}^1
A[s]^\dagger\rho A[s],
\end{eqnarray*}
which works as  
\begin{eqnarray*}
\mathcal{E}(|0\rangle\langle0|)
&=&\cos^2\frac{\theta}{2}|0\rangle\langle0|
+\sin^2\frac{\theta}{2}|1\rangle\langle1|\\
\mathcal{E}(|1\rangle\langle1|)
&=&
\sin^2\frac{\theta}{2}|0\rangle\langle0|
+\cos^2\frac{\theta}{2}|1\rangle\langle1|.
\end{eqnarray*}
Under the correspondence
\begin{eqnarray*}
|0\rangle\langle0| &\iff& 
\left(
\begin{matrix}
1\\
0
\end{matrix}
\right)\\
|1\rangle\langle1| &\iff& 
\left(
\begin{matrix}
0\\
1
\end{matrix}
\right),
\end{eqnarray*}
the map $\mathcal{E}$ works as 
\begin{eqnarray*}
\mathcal{E}=
\left(
\begin{matrix}
\cos^2\frac{\theta}{2}&\sin^2\frac{\theta}{2}\\
\sin^2\frac{\theta}{2}&\cos^2\frac{\theta}{2}.
\end{matrix}
\right).
\end{eqnarray*}
Since this $2\times2$ matrix
is diagonalized by the unitary
\begin{eqnarray*}
\frac{1}{\sqrt{2}}\left(
\begin{matrix}
1&1\\
1&-1
\end{matrix}
\right)
\end{eqnarray*}
as diag($1,\cos\theta$), 
we obtain
\begin{eqnarray*}
\langle-|\Big[
\mathcal{E}\circ...\circ\mathcal{E}(|0\rangle\langle0|)\Big]|+\rangle
=\frac{1}{2}\cos^N\theta.
\end{eqnarray*}
Since the norm of $\Phi(|+\rangle)_1^N$
and $\Phi(|-\rangle)_1^N$ is $\frac{1}{\sqrt{2}}$,
the normalized inner product is $\cos^N\theta$.


{\bf Appendix 4: Gate teleportation}.---
Let us first consider the teleportation of a single-qubit 
unitary~\cite{teleQC}.
Without loss of generality, we can start with
\begin{eqnarray*}
\sum_{s=0}^1\Phi(|\phi_s\rangle)_2^N\otimes|m_s\rangle_1,
\end{eqnarray*}
since if $A[0]$ and $A[1]$ have the form Eq.~(\ref{Asimple}), we immediately
obtain this state, and
if $A[0]$ and $A[1]$ have the form Eq.~(\ref{Acomplicated}), we 
obtain this state by using the filtering.
By the definition of the computational quantum wire, the basis change
$\{|\phi_0\rangle,|\phi_1\rangle\}\to\{|0\rangle,|1\rangle\}$
is always possible:
\begin{eqnarray*}
\sum_{s=0}^1\Phi(|s\rangle)_k^N\otimes|m_s\rangle_1
\end{eqnarray*}
for certain $k$.
Then, by projecting the zeroth and first qubit of
\begin{eqnarray*}
\Big[\sum_{s=0}^1\Phi(|s\rangle)_k^N\otimes|m_s\rangle_1\Big]\otimes|\psi\rangle_0
\end{eqnarray*}
onto 
$(I_1\otimes U^\dagger_0)|B\rangle_{10}$, 
where $I_1$ is the identity operator on the first qubit,
$U_0$ is a unitary on the zeroth qubit,
and $|B\rangle_{10}$ is one of the Bell basis
\begin{eqnarray*}
&&|m_0\rangle_1\otimes|0\rangle_0+|m_1\rangle_1\otimes|1\rangle_0,\\
&&|m_0\rangle_1\otimes|0\rangle_0-|m_1\rangle_1\otimes|1\rangle_0,\\
&&|m_0\rangle_1\otimes|1\rangle_0+|m_1\rangle_1\otimes|0\rangle_0,\\
&&|m_0\rangle_1\otimes|1\rangle_0-|m_1\rangle_1\otimes|0\rangle_0,
\end{eqnarray*}
we obtain the desired state $\Phi(U|\psi\rangle)$ up to the bit
or phase errors in the correlation space.

\begin{figure}[htbp]
\begin{center}
\includegraphics[width=0.13\textwidth]{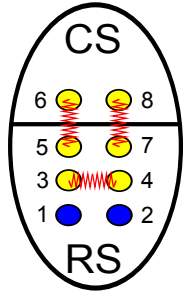}
\end{center}
\caption{(Color online.) 
Teleportation of the controlled-Z gate from the real space to
the correlation space. Two blue circles represent the two-qubit
state $|\psi\rangle$. The Red zigzag line between
the third and fourth qubits represents the real physical entanglement.
Other red zigzag lines represent the virtual "entanglement".
The first, third, and fifth qubits are projected. 
The second, fourth, and seventh qubits are projected.
Then, the state of the sixth and eighth qubits is 
$CZ|\psi\rangle$ up to some trivial errors.
} 
\label{teleCZ}
\end{figure}

Next, let us consider the teleportation of the controlled-Z 
gate~\cite{teleQC,RVB}
\begin{eqnarray*}
|0\rangle\langle0|\otimes I+
|1\rangle\langle1|\otimes Z.
\end{eqnarray*}
Without loss of generality, 
we can start with the state (see Fig.~\ref{teleCZ})
\begin{eqnarray*}
&&\Big[\sum_{s=0}^1\Phi(|\phi_s\rangle_6)\otimes|m_s\rangle_5\Big]
\otimes\Big[\sum_{r=0}^1\Phi(|\phi_r\rangle_8)\otimes|m_r\rangle_7\Big]\\
&&\otimes|B\rangle_{34}
\otimes|\psi\rangle_{12},
\end{eqnarray*}
where 
\begin{eqnarray*}
|\psi\rangle_{12}&=&c_{00}|0\rangle_1\otimes|0\rangle_2
+c_{01}|0\rangle_1\otimes|1\rangle_2\\
&&+c_{10}|1\rangle_1\otimes|0\rangle_2
+c_{11}|1\rangle_1\otimes|1\rangle_2
\end{eqnarray*}
is the state of the first and second qubits,
and
\begin{eqnarray*}
|B\rangle_{34}\equiv|0\rangle_3\otimes|0\rangle_4
+|1\rangle_3\otimes|1\rangle_4
\end{eqnarray*}
is the maximally entangled state of the third and fourth qubits.
The first, third, and fifth qubits are projected onto
\begin{eqnarray*}
|0\rangle_1\otimes|0\rangle_3\otimes|m_0\rangle_5
+|1\rangle_1\otimes|1\rangle_3\otimes|m_1\rangle_5,
\end{eqnarray*}
whereas
the second, fourth, and seventh qubits are projected onto
\begin{eqnarray*}
|0\rangle_2\otimes|+\rangle_4\otimes|m_0\rangle_7
+|1\rangle_2\otimes|-\rangle_4\otimes|m_1\rangle_7.
\end{eqnarray*}
Then, we obtain the desired state
\begin{eqnarray*}
&&c_{00}\Phi(|\phi_0\rangle_6)\otimes\Phi(|\phi_0\rangle_8)
+c_{01}\Phi(|\phi_0\rangle_6)\otimes\Phi(|\phi_1\rangle_8)\\
&&+c_{10}\Phi(|\phi_1\rangle_6)\otimes\Phi(|\phi_0\rangle_8)
-c_{11}\Phi(|\phi_1\rangle_6)\otimes\Phi(|\phi_1\rangle_8).
\end{eqnarray*}
We obtain the same result up to the trivial errors if the six qubits
are projected onto other states.

{\bf Appendix 5: Entanglement swapping}.---
Let us see Fig.~\ref{teleportation} (c).
The initial state is
\begin{eqnarray*}
\Big[\sum_{s=0}^1\Phi(|\phi_s\rangle)\otimes|m_s\rangle\Big]
\otimes
\Big[\sum_{r=0}^1\Phi(|\phi_r\rangle)\otimes|m_r\rangle\Big].
\end{eqnarray*}
Two real physical qubits are projected onto one of the Bell basis
\begin{eqnarray*}
&&|m_0\rangle\otimes|m_0\rangle+|m_1\rangle\otimes|m_1\rangle\\
&&|m_0\rangle\otimes|m_0\rangle-|m_1\rangle\otimes|m_1\rangle\\
&&|m_0\rangle\otimes|m_1\rangle+|m_1\rangle\otimes|m_0\rangle\\
&&|m_0\rangle\otimes|m_1\rangle-|m_1\rangle\otimes|m_0\rangle.
\end{eqnarray*}
Then,
we obtain the desired state
\begin{eqnarray*}
\sum_{s=0}^1\Phi(|\phi_s\rangle)
\otimes
\Phi(|\phi_s\rangle)
\end{eqnarray*}
up to the phase or the bit errors.





\begin{thebibliography}{00}
\bibitem{one-way}
R. Raussendorf and H. J. Briegel, Phys. Rev. Lett. {\bf86}, 5188 (2001).

\bibitem{Gross}
D. Gross and J. Eisert, Phys. Rev. Lett. {\bf98}, 220503 (2007).


\bibitem{Cai}
J. M. Cai, et. al.,
Phys. Rev. Lett. {\bf103}, 050503 (2009).
\bibitem{power1}
M. Van den Nest, et. al.,
Phys. Rev. Lett. {\bf97}, 150504 (2006).
\bibitem{teleQC}
D. Gottesman and I. L. Chuang, Nature {\bf402}, 390 (1999).

\bibitem{RVB}
F. Verstraete and J. I. Cirac, Phys. Rev. A {\bf70}, 060302(R) (2004).
\bibitem{clone}
W. K. Wootters and W. H. Zurek, Nature {\bf299}, 802 (1982).
\bibitem{tele}
C. H. Bennett, et. al.,
Phys. Rev. Lett. {\bf70}, 1895 (1993).
\bibitem{dense}
C. H. Bennett and S. J. Wiesner, Phys. Rev. Lett. {\bf69}, 2881 (1992).
\bibitem{repeater}
H. J. Briegel, et. al.,
Phys. Rev. Lett. {\bf81}, 5932 (1998).

\bibitem{Furusawa}
F. Furusawa, et. al.,
Science {\bf282}, 706 (1998);
M. A. Nielsen, et. al.,
Nature {\bf396}, 52 (1998);
M. Riebe, et. al., Nature {\bf429}, 734 (2004);
M. D. Barrett, et. al., Nature {\bf429}, 737 (2004);
J. H. Reina and N. F. Johnson, Phys. Rev. A {\bf63}, 012303 (2000);
J. Sherson, et. al., Nature {\bf443}, 557 (2006);
Y. A. Chen, et. al., Nature Physics {\bf4}, 103 (2008).
\bibitem{experiment}
W. B. Gao, et. al., arXiv:1004.4162.
\bibitem{Yip}
S. K. Yip, Phys. Rev. Lett. {\bf90}, 250402 (2003).
\bibitem{Brennen}
G. K. Brennen, et. al., New J. Phys. {\bf9}, 138 (2007).
\bibitem{MiyakeAKLT}
G. K. Brennen and A. Miyake, Phys. Rev. Lett. {\bf101}, 010502 (2008).
\bibitem{KLM}
E. Knill, et. al., Nature {\bf46}, 409 (2001).
\end{thebibliography}
\end{document}